\documentclass[aps,twocolumn,showpacs,longbibliography]{revtex4-1}
\usepackage{bm}
\usepackage{color} 
\usepackage{graphicx}
\usepackage{epsfig}
\usepackage{hyperref}
\usepackage{natbib}
\usepackage{amsmath}
\usepackage{amssymb}

\newcommand{\rmi}{{\rm i}}

\newcommand {\e}{{\rm e}}

\renewcommand {\i}{{\rm i}}

\newcommand{\av}[1]{\left\langle #1\right\rangle}

\begin{document}

\title{Excitation of spin density and current by coherent light pulses in QWs}

\author{A.\,V.\,Poshakinskiy}
\author{S.\,A.\,Tarasenko}
\affiliation{A.\,F.\,Ioffe Physical-Technical Institute, 194021 St.~Petersburg, Russia}

\begin{abstract}
We study the orbital and spin dynamics of charge carriers induced by non-overlapping linearly polarized light pulses in semiconductor quantum wells (QWs). It is shown that such an optical excitation with coherent pulses leads to a spin orientation of photocarriers and an electric current. The effects are caused by the interference of optical transitions driven by individual pulses. The distribution of carriers in the spin and momentum spaces depends on the QW crystallographic orientation and can be efficiently controlled by the pulse polarizations, time delay and phase shift between the pulses, as well as an external magnetic field.
\end{abstract}


%
%
%
%
%
%
%
%

\maketitle

{\bf Introduction.} The control of quantum states in nanostructures by ultrashort light pulses is at the heart of modern solid-state optics. By applying a sequence of coherent light pulses with defined relative phases and polarizations, one can efficiently manipulate the
quantum dynamics provided the excited system stays coherent for a sufficiently long time. Such a coherent control has been demonstrated for exciton population and polarization in QWs~\cite{Heberle1995,Marie1997,Gridnev2011}, trions in QWs~\cite{Versluis2010,Langer2012},
excitions and charge carriers in quantum dots~\cite{Toda2000,Htoon2002,Godden2012}, polaritons in semiconductor microcavities~\cite{Renucci2003}, see also Refs.~\cite{Cundiff2008,Ramsay2010} for recent surveys. It has been also shown that the coherent light pulses can cause a real-space shift of electronic charges in semiconductors~\cite{Priyadarshi2012,Priyadarshi2013}.
Previous research was focused on optical transitions beweet discrete levels. Here, stimulated by progress in ultrafast optical spectroscopy, we present the theoretical study of orbital and spin dynamics of free carriers in quantum wells (QWs) induced by non-overlapping linearly-polarized light pulses. We show that the interference of optical transitions caused by individual pulses leads to a spin polarization of photoelectrons and to an electric current. The spin polarization as well as the photocurrent direction and magnitudes are determined by pulse characteristics and QW crystallographic orientation.

\begin{figure}[b!]
  \includegraphics[width=0.75\columnwidth]{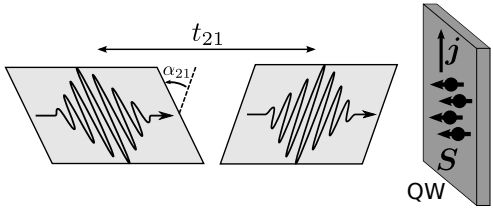}
\caption{The sketch of the QW excitation by two coherent linearly polarized light pulses. The excitation leads to a spin polarization of carriers and an electric current.}
\label{fig1}
\end{figure}

{\bf Model.} We consider excitation of a QW by the sequence of two optical pulses which are not overlapped with each other, see Fig.~1. The pulses are assumed to be linearly polarized and normally incident upon the structure. The electromagnetic fields of the pulses are described by the vector potentials
\begin{eqnarray}
\bm{A}_1(t) = \bm{e}_1 a_1(t) \e^{-\i\omega t} +  c.c. \:, \hspace{2.1cm} \\
\bm{A}_2(t) = \bm{e}_2 a_2(t-t_{21}) \e^{-\i\omega (t-t_{21}) - i \phi} + c.c. \:, \nonumber
\end{eqnarray}
where $\bm{e}_1$ and $\bm{e}_1$ are the unit polarization vectors of the pulses, $a_1(t)$ and $a_2(t-t_{21})$ are the pulse envelopes centered at $t=0$ and $t=t_{21}$, respectively, $t_{21}$ is the time delay between the pulses, $\omega$ is the carrier frequency tuned to the QW bandgap, and $\phi$ is the phase shift.

The pulses cause direct optical transitions between the ground heavy hole $hh1$ and electron $e1$ subbands. The first pulse acting on the equilibrium electron system, where all valence-band states are occupied while conduction-band states are empty, induces the optical transitions and thereby creates an {\it interband} polarization in the QW. Such an interband polarization oscillates at optical frequencies and interferes with the interband polarization induced by the second pulse. The interference occurs provided the electron system stores the interband coherence between the light pulses. Thus, the final distribution of electrons in the conduction band is determined not only by the individual pulse absorption but also by the interference processes despite the fact that the pulses are separated in time. 

We calculate the electron and hole distributions in the the momentum and spin spaces by solving the quantum kinetic equation for the density matrix~\cite{Gridnev2011}. In this approach, the electron system in both conduction and valence subbands is described by the density matrix 
\begin{equation}
\rho = 
\left(
\begin{array}{cc} 
\rho_{cc} & \rho_{cv} \\
\rho_{vc} & \rho_{vv} 
\end{array}
\right) 
\end{equation}
which consists of four $2\times2$ blocks. The diagonal blocks $\rho_{cc}$ and $\rho_{vv}$ represent the spin and pseudospin density matrices for the $e1$ and $hh1$ subbands, respectively. The off-diagonal blocks $\rho_{cv}$ and $\rho_{vc}$ describe correlations between the conduction-band and valence-band states. The density matrix $\rho$ satisfies the quantum kinetic equation
\begin{equation}\label{rho_gen}
\frac{\partial \rho}{\partial t} = -\frac{\rmi}{\hbar} [H,\rho] + {\rm St} \rho \:,
\end{equation}
where $H$ is the Hamiltonian,
\begin{equation}
H = 
\left(
\begin{array}{cc} 
H_{e1} & V_{cv} \\
V_{vc} & H_{hh1} 
\end{array}
\right) ,
\end{equation}
its diagonal blocks $H_{e1}$ and $H_{hh1}$ determine the spectra in the electron and hole subbands, $V_{cv}$ and $V_{vc}$ are the matrices
of electron-photon interaction operator, and ${\rm St} \rho$ is the collision integral describing relaxation processes due to electron and hole scattering by defects, phonons, etc. 
In the canonical basis of the conduction-band and valence-band states with the spin projections $\pm 1/2$ and $\pm 3/2$, respectively, along the quantization axis $z$, the matrices of the interaction operator have the form~\cite{Ivchenko_book}
\begin{equation}
V_{cv} = \frac{e J P_{cv}}{\sqrt{2} \, m_0 c} 
\left[ 
\begin{array}{cc} 
A_x + \rmi A_y & 0 \\
0 & -A_x + \rmi A_y 
\end{array}
\right]
\end{equation}
and $V_{vc}=V_{cv}^\dag$, where $e$ is the electron charge, $m_0$ is the free electron mass, $c$ is the speed of light, $J$ is the overlap integral of the electron and hole envelope functions, and $P_{cv} = \langle S | p_x | X \rangle$ is the interband matrix element of the momentum operator. We assume that the duration of each optical pulse is much shorter than any relaxation time in the system and, therefore, neglect the last term in Eq.~\eqref{rho_gen} in calculating the density matrix evolution within the pulse action. Between the pulses, the electron system may relax which is taken into account in the relaxation time approximation: the anisotropic parts of the conduction-band $\rho_{cc}$ and valence-band $\rho_{vv}$ density matrices decay with the electron and hole relaxation times, $\tau_e$ and $\tau_h$, respectively. The off-diagonal components $\rho_{cv}$ and $\rho_{vc}$ are destroyed by any scattering process and, under quite general assumptions, decay at the rate  $\gamma = (1/\tau_e + 1/\tau_h)/2$. Below we solve the quantum kinetic equation and discuss the results of calculations for different QW systems.   

{\bf Coherent optical orientation.} First, we consider the simple case where the electron dispersions in the $e1$ and $hh1$ subbands are parabolic and spin-degenerate
\begin{equation}
H_{e1} = \frac{\hbar^2 \bm k^2}{2 m_e} \:, \;\; H_{hh1} = - E_g - \frac{\hbar^2 \bm k^2}{2 m_h} \:. 
\end{equation}
Here, $\bm k$ is the wave vector, $m_e$ and $m_h$ are the in-plane effective masses in the $e1$ and $hh1$ subbands, and $E_g$ is the effective bandgap in the QW. 
The solution of Eq.~\eqref{rho_gen} to the second order in the electromagnetic field amplitude shows that the spin density matrix in the $e1$ subband after the pulses has the form
\begin{align}\label{rhocc}
&\rho_{cc} = R 
\Big\{ E_1 E_1^\dag \, |a_1(\omega_{\bm k}-\omega)|^2 + E_2 E_2^\dag \,|a_2(\omega_{\bm k}-\omega)|^2  \\
& + [ E_1 E_2^\dag \, a_1(\omega_{\bm k}-\omega) a_2^*(\omega_{\bm k}-\omega) \e^{-\rmi \Phi_{\bm k}} + h.c.] \, \e^{-\gamma t_{21}} \Big\} \nonumber , 
\end{align}
where $R = (1/2) [e J |P_{cv}| / (\hbar m_0 c)]^2$, $E_j = e_{j,x} \sigma_z + i e_{j,y}I$ are the matrices determined by pulse polarizations ($j=1,2$), $\sigma_z$ is the Pauli matrix, $I$ is the $2\times2$ unit matrix, 
$a_j(\omega) = \int a_j(t) \e^{\rmi \omega t} d t$ are the Fourier components of the pulse envelopes, $\Phi_{\bm k}=\omega_{\bm k} t_{21}-\varphi_{21}$, $\omega_{\bm k} = [\hbar^2 \bm k^2 /(2\mu) + E_g]/\hbar$, $\mu=m_e m_h/(m_e + m_h)$ is the reduced mass, and $h.c.$ stands for the Hermitian conjugate term. The first and second terms on the right-hand side of Eq.~\eqref{rhocc} describe the electron distribution in the $e1$ subband that would be created by the first and second optical pulses if they were independent, while the last term comes from interference. 

The spin density matrix known, one can readily find the particle $f_{\bm k}$ and spin $\bm s_{\bm k}$ distribution functions, $ f_{\bm k} = {\rm Tr} \rho_{cc}$, $\bm s_{\bm k} = {\rm Tr} (\bm \sigma \rho_{cc})/2$. For the particular case of linearly-polarized pulses with the same envelope, the distribution functions have the form 
\begin{align}
&f_{\bm k} = f_0 \left( 1+ \cos\Phi_{\bm k} \cos \alpha_{21} \,\e^{-\gamma t_{21}}  \right)\:, \label{fk}\\
&s_{\bm k,z}= - \frac{f_0}{2} \sin\Phi_{\bm k} \sin \alpha_{21} \, \e^{-\gamma t_{21}} \:,\label{Sk}
\end{align}
where $f_0 = 4 R |a(\omega_{\bm k} - \omega)|^2$ is the distribution function created by incoherent pulses, $a(t) \equiv a_1(t) = a_2(t)$ and $\alpha_{21}$ is the angle between pulse polarization planes, $\cos \alpha_{21} = \bm e_1 \cdot \bm e_2$, $\sin \alpha_{21} = [\bm e_1 \times \bm e_2]_z$. 
\begin{figure}[t]
  \includegraphics[width=0.95\columnwidth]{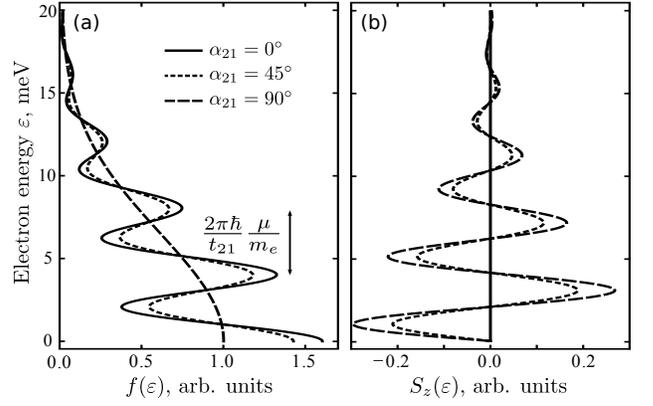}
\caption{Energy distributions of (a) electron population and (b) electron spin created by two coherent linearly polarized pulses.
Calculations are presented for the optical frequency $\omega = E_g/\hbar$, the Gaussian pulse profile $a(t)=a_0 \exp [-t^2/(2\tau^2)]$, $\tau=0.1\,$ps, $t_{21}=1\,$ps, $\gamma = 0.5$\,ps$^{-1}$, $\Phi_{\bm k=0}=0$, and $m_h \gg m_e$.}
\label{fig2}
\end{figure}

Figure~\ref{fig2}a shows the energy distribution of photoelectrons in the $e1$ subband immediately after the QW excitation. The curves are calculated after Eq.~\eqref{fk} for different angles between the pulse polarization planes. The distribution function scale is determined by the spectral width of the pulses. Besides, the function contains pronounced oscillations which originate from the pulse interference. The energy distance between the adjacent maxima is given by $\Delta\varepsilon = (2\pi\hbar/t_{21})(\mu/m_e)$. The oscillation amplitude is proportional to $\cos \alpha_{21}$: The oscillations are maximal for co-polarized pulses and vanish for cross-polarized pulses.

More interestingly, the interference of optical transitions leads to the spin orientation of electrons along the $z$ axis despite the fact that both light pulses carry no angular momentum and an optical orientation by any individual pulse does not occur~\cite{Marie1997}.
The efficiency of such a coherent optical orientation can be high 
provided the interband decoherence is slow enough. Figure~\ref{fig2}b shows the energy distribution of the spin component $S_z$ in the $e1$ subband calculated after Eq.~\eqref{Sk} for different angles $\alpha_{21}$. The spin polarization oscillates as a function of the electron energy with the same frequency as does the electron density. The degree of spin polarization reaches a maximum for cross-polarized pulses, the polarization has the opposite signs for positive and negative $\alpha_{21}$ angles and vanishes for co-polarized light pulses.

Being caused by quantum interference, both the population oscillations and spin orientation are highly sensitive to the pulse phase shift $\varphi_{21}$ and the time delay $t_{21}$. The latter can be used in experiments to study the electron interband  coherence and dephasing.

{\bf Magnetic field effect.} Now, we analyze the influence of an external magnetic field $\bm{B}$ on the coherent optical orientation.
Generally, the magnetic field leads to the cyclotron motion of charge carriers and Zeeman splitting of the spin states. With allowance for the both effects, the effective Hamiltonians $H_{e1}$ and $H_{hh1}$ are given by
\begin{equation}\label{H_e1}
H_{e1} = \frac{1}{2m_e} \left( \bm{p} - \frac{e}{c} \bm{A}_B \right)^2 + \frac{\hbar}{2} \bm{\sigma} \cdot \bm{\Omega}_{e} \:, 
\end{equation}
\[
H_{hh1} = - \left[ E_g + \frac{1}{2m_h} \left( \bm{p} - \frac{e}{c} \bm{A}_B \right)^2 \right] + 
\frac{\hbar}{2} \bm{\sigma} \cdot \bm{\Omega}_{h}  \:, 
\]
where $\bm{p}$ is the momentum operator, $\bm{A}_B$ is the vector potential corresponding to the magnetic field $\bm{B}$, 
$\bm{\Omega}_{e}$ and $\bm{\Omega}_{h}$ are the Larmor frequencies corresponding to the $e1$ and $hh1$ subband spin splitting, respectively, $\Omega_{e,\alpha} = (\mu_0 /\hbar) g_{\alpha \beta}^{(e1)} B_{\beta}$ and $\Omega_{h,\alpha} = (\mu_0 /\hbar) g_{\alpha \beta}^{(hh1)} B_{\beta}$, $\mu_0$ is the Bohr magneton, and $g_{\alpha \beta}^{(e1)}$ and $g_{\alpha \beta}^{(hh1)}$ are the effective $g$-factor tensors. 

We assume that the magnetic field is weak enough and threat it quasi-classically. We also consider that the cyclotron and Larmor frequencies are small compared to the inverse pulse duration and, therefore, neglect the magnetic field influence on the optical transitions. Between the pulses the time evolution of the interband component of the density matrix is described by the equation
\begin{align}\label{kin_cv}
\frac{\partial \rho_{cv}}{\partial t} = - \frac{\bm{v}_{e}+\bm{v}_{h}}{2} \cdot \frac{\partial \rho_{cv}}{\partial \bm{r}}
 - \frac{e (\bm{v}_e + \bm{v}_h) \times \bm{B}}{2c \hbar} \cdot \frac{\partial \rho_{cv}}{\partial \bm{k}}  \\
- \frac{\rmi}{2} \left( \bm{\sigma} \cdot \bm{\Omega}_e \, \rho_{cv} - \rho_{cv} \, \bm{\sigma} \cdot \bm{\Omega}_h \right) - (\rmi \omega_{\bm{k}}+\gamma) \rho_{cv} \:,\nonumber
\end{align}
where $\bm{v}_e = \hbar \bm{k}/m_e$ and $\bm{v}_h = - \hbar \bm{k}/m_h$. Equation~\eqref{kin_cv} can be derived by considering the Wigner quasi-probability distribution~\cite{Akhiezer_book}. It is similar to the kinetic equation for the intraband spin density matrix. For the case under study, where $\rho_{cv}$ is spatial homogenous and independent of the wave vector direction, the first and second terms on the right-hand side of Eq.~\eqref{kin_cv} vanish and only the Larmor precession affects the spin dynamics.

The calculation of the quantum kinetic equation shows that the spin density matrix of photoelectrons immediately after the light pulses
has the form 
\begin{align}\label{rhocc_B}
 \rho_{cc} &= \frac{f_0}{4}\left[ 1 + \exp\left(-\rmi \frac{\bm\sigma \cdot \bm\Omega_e  t_{21}}{2} \right) E_1 \exp\left(\rmi \frac{\bm\sigma \cdot \bm\Omega_h  t_{21}}{2} \right) \right. \nonumber \\
 & \hspace{3cm}\times \left.  E_2^\dag \, \e^{-\rmi \Phi_{\bm k} -\gamma t_{21}} \right] + h.c. 
\end{align}
If both $\bm\Omega_e$ and $\bm\Omega_h$ are parallel to the QW normal $z$, which is usually realized by applying the out-of-plane magnetic field, the magnetic field introduces the phase shift $(\bm\Omega_{e} - \bm\Omega_{h})t_{21}$. In this case, the particle and spin distribution functions are described by Eqs.~(\ref{fk})-(\ref{Sk}) where $\alpha_{21}$ is replaced by $\alpha_{21}+(\bm\Omega_{e,z} - \bm\Omega_{h,z})t_{21}/2$. The optical orientation occurs now even for co-polarized pulses, $\alpha_{21}=0$, and is efficient if the frequency $|\bm\Omega_{e} - \bm\Omega_{h}|$ is comparable to $1/t_{21}$. Estimations show that such a condition is achievable for real semiconductor QWs: For the $g$-factor difference $|g_{zz}^{hh1} - g_{zz}^{e1}| = 2$, the magnetic field $B = 2\,$T, and the time delay $t_{21} = 1 \,$ps, one obtains $|\bm\Omega_{e} - \bm\Omega_{h}| \tau_{21} \approx 0.4$.       

For arbitrary direction of the magnetic field, the spin distribution induced by two identical co-polarized pulses is given by
\begin{align}\label{Sk_B}
&\bm s_{\bm k}=\frac{f_0}{2}  \sin\Phi_{\bm k}\, \e^{-\gamma t_{21}} \Bigg( \frac{\bm\Omega_e\times\bm{\tilde\Omega}_h}{\Omega_e\,\Omega_h}  \sin\frac{\Omega_e t_{21}}{2} \sin\frac{\Omega_h t_{21}}{2} \nonumber \\ 
& - \frac{\bm\Omega_e}{\Omega_e}  \sin\frac{\Omega_e t_{21}}{2} \cos\frac{\Omega_h t_{21}}{2}+
\frac{\bm{\tilde\Omega}_h}{\Omega_h}  \cos\frac{\Omega_e t_{21}}{2} \sin\frac{\Omega_h t_{21}}{2}
\Bigg) ,
\end{align}
where $\bm{\tilde\Omega}_h$ is the vector obtained from $\bm\Omega_h$ by rotating around the $z$ axis by the $\pi-2\alpha$, with
$\alpha = \arctan e_y/e_x$ being the angle between the polarization vector $\bm{e}$ and the $x$ axis, i.e., 
 $\tilde\Omega_x = -\Omega_x\cos2\alpha+\Omega_y\sin2\alpha$, $\tilde\Omega_y = -\Omega_x\sin2\alpha-\Omega_y\cos2\alpha$, and $\tilde\Omega_z = \Omega_z$. Generally, the spin orientation is not colinear to the vector $\bm \Omega_e$ and, therefore, will precess 
around $\bm \Omega_e$ after the pulses. Such a polarization can considerably exceed the equilibrium thermal polarization of electrons in the magnetic field.

\begin{figure}
  \includegraphics[width=0.75\columnwidth]{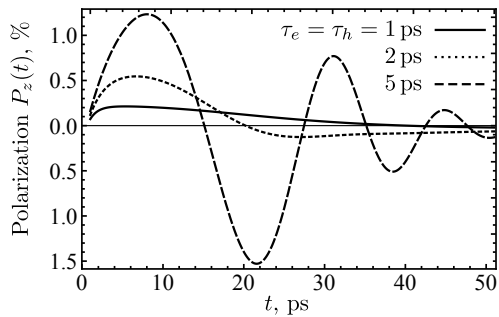}
\caption{Time dependence of the spin polarization $P_z(t)$ generated by two co-plarized pulses in asymmetric (001)-grown QW.  
Curves are plotted after Eq.~\eqref{Szt} for $\beta_e=\beta_h=30$\,meV$\cdot$\AA , $\alpha=0$, $\Phi_{\bm k=0}=\pi/2$ and other pulse parameters given in the caption of Fig.~\ref{fig2}.}
\label{fig3}
\end{figure}
{\bf Effect of spin-orbit interaction.} Equations~\eqref{rhocc_B} and~\eqref{Sk_B} obtained for the Zeeman splitting caused by an external magnetic field are also valid for the spin splitting of the $e1$ and $hh1$ subbands induced by Rashba/Dresselhaus spin-orbit coupling. It indicates that the optical orientation by co-polarized pulses can occur in QWs even in the absence of external magnetic field. Similar effect for a single-pulse excitation with linearly polarized radiation was considered in Refs.~\cite{tarasenko2005,Gorelov2011}. In the case of spin-orbit splitting, the effective Larmor frequencies $\bm\Omega_e$ and $\bm\Omega_h$ depend on the wave vector $\bm{k}$ and are odd functions of $\bm{k}$~\cite{Winkler_book}. The particular form of $\bm\Omega_e(\bm k)$ and $\bm\Omega_h(\bm k)$ is determined by the QW crystallographic orientation and structure design. Immediately after the excitation with co-polarized light pulses, the spin distribution in the $\bm k$ space is given by Eq.~\eqref{Sk_B} and contains both odd-in-$\bm k$ and even-in-$\bm k$ terms. Therefore, the average spin polarization of the electron gas is non-zero. After the pulses, the electron spin dynamics in the effective magnetic field is described by the equation
\begin{equation}\label{kin_sk}
\frac{\partial\bm{s_k}}{\partial t} + \bm{s_k} \times \bm\Omega_{e}= -\frac{s_{\bm k} - \av{s_{\bm k}}}{\tau_e} \:,
\end{equation}
where the the angular brackets denote averaging over the directions of $\bm k$. Equation~\eqref{kin_sk} follows from Eq.~\eqref{rho_gen} and is valid in the approximation of elastic electron scattering for the times shorter than the energy relaxation time. The time evolution of the total electron spin $\bm S(t)=\sum_{\bm k} \bm{s_k}(t)$ after the pulses ($t>t_{21}$) can be calculated by solving Eq.~\eqref{kin_sk} with the initial condition Eq.~\eqref{Sk_B}.
The dependence $\bm S(t)$ is determined by particular form of spin-orbit interaction.

As an example, we elaborate this effect for a QW with the $(001)$ crystallographic orientation. In such structures the effective magnetic fields in both $e1$ and $hh1$ subbands lie in the QW plane and, microscopically, can be caused by bulk inversion asymmetry (Dresselhaus effect) or structure inversion asymmetry (Rashba effect)~\cite{Winkler_book}. The symmetry analysis shows that the optical orientation by co-polarized pulses may occur in structures of the $C_{2v}$ point group (asymmetric QWs) and is forbidden in systems of the $D_{2d}$ group (symmetric QWs) and $C_{\infty v}$ group (uniaxial approximation). Therefore, to obtain such an optical orientation one should take into account both Rashba and Dresselhaus contributions to the spin-orbit splitting. We assume for simplicity that the spin-orbit splitting of the $e1$ subband is determined by the Rashba term $\bm\Omega_{e}=2\beta_e/\hbar\, (k_y,-k_y,0)$ while the splitting of the $hh1$ subbband is given by the Dresselhaus term $\bm\Omega_{h}=2\beta_h/\hbar\, (k_x,k_y,0)$~\cite{Rashba1988,Durnev2014}, where $x\parallel[100]$, $y\parallel[010]$ and $z\parallel[001]$ are the cubic axes. In this particular case, the value of spin-orbit splitting $\hbar\Omega_e$ is independent of the wave vector direction, which enables analytical solution of Eq.~\eqref{kin_sk}~\cite{Gridnev01,Poshakinskiy11a}. The straightforward calculation shows that the time dependence of the total electron spin at $t>t_{21}$ has the form
\begin{align}\label{Szt}
&S_z (t)=- \cos 2\alpha \sum\limits_{\bm k}\frac{f_0}{2}  \sin\Phi_{\bm k} \sin \frac{\beta_hk t_{21}}{\hbar}\,  \\\nonumber&
\times \Bigg\{  \left[\sin \frac{\beta_e k t_{21}}{\hbar} + \frac{4 \beta_e k \tau_{e}}{\hbar} \cos \frac{\beta_e k t_{21}}{\hbar}\right] \frac{\sin \nu (t-t_{21})}{2 \nu \tau_{e}} \\\nonumber&
 +\sin \frac{\beta_e k t_{21}}{\hbar} \cos \nu (t-t_{21}) \Bigg\} \,\e^{-(t-t_{21})/2\tau_{e}}\, \e^{-\gamma t_{21}} \:,
\end{align}
where $\nu=\sqrt{(2\beta_e k/\hbar)^2 - 1/(4\tau_{e}^2)}$. Figure~\ref{fig3} shows the time dependence of the total spin polarization $P_z(t)=2S_z(t)/\sum_{\bm k} f_{\bm k}$ generated by two co-polarized pulses. The curves are plotted after Eq.~\eqref{Szt} for different scattering times $\tau_e=\tau_h=1/\gamma$. 
Just after the second pulse arrives ($t=t_{21}$), the electron gas gains a small spin polarization caused by the interference of the optical transitions induced by the pulses. This polarization is described by the first term in the right-hand side of Eq.~\eqref{Sk_B} and is proportional to the product $\bm \Omega_e \times \bm \Omega_h$. Besides the average spin polarization, the pulses create an asymmetric contribution to the spin distribution function $\bm S_{\bm k}$ given by the last two terms in Eq.~\eqref{Sk_B}. The subsequent kinetics of the asymmetric spin distribution in the effective magnetic field leads to an additional contribution to $S_z$ giving rise to an increase in the net spin polarization, see Fig.~\ref{fig3}. At even larger times, the spin polarization monotonously decays or exhibits 
damping oscillations at the frequency $\Omega_e$ depending on the ratio between the spin precession frequency $\Omega_e$ and the electron collision rate $1/\tau_e$, cf. solid and dashed curves in Fig.~\ref{fig3}.

{\bf Photogalvanic effect.} The $\bm k$-linear spin-orbit splitting of the electron or hole subbands can lead to the generation of a  photocurrent which is sensitive to the radiation helicity and reverses its direction upon switching the circular polarization sign. Microscopically, the current is caused by an asymmetric distribution of photoexited carriers in the momentum space due spin-dependent selection rules for optical transitions. Such photogalvanic effects in QW structures are actively studied both experimentally and theoretically, see recent surveys~\cite{Ivchenko2008,Rioux2012,Ganichev2014}. Below we demonstrate that the photocurrent can be excited by two coherent linearly polarized pulses. 

At the normal incidence of radiation, the photogalvanic effect is allowed in QWs of sufficiently low spatial symmetry only~\cite{Shalygin2007}, where the effective magnetic field has an out-of-plane component. We consider the effect for symmetric QWs with the  $(110)$ crystallographic orientation. In such  structures, the spin-orbit splittings of the $e1$ and $hh1$ subbands have the form $\bm\Omega_e = 2\beta_e /\hbar\: (0,0,k_{x'})$ and $\bm\Omega_h = 2\beta_h /\hbar\: (0,0,k_{x'})$, respectively, where $x'\parallel [1\bar 10]$ and $y'\parallel [00\bar 1]$ are the in-plane axes, and $z'\parallel [110]$ is the QW normal. The splitting is linear in $k_{x'}$ and leads to the electric current along the $x'$ axis. The energy distribution of the photocurrent density in the $e1$ subband is given by
\begin{equation}\label{current0}
j_{x'}(\varepsilon) =e \sum\limits_{\bm k,s}  v_{s,x'} f_{\bm k,s}
\,\delta\left( \varepsilon_{\bm k,s} - \varepsilon \right) \:,
\end{equation}
where $s=\pm1/2$ is electron spin projection onto the $z'$ axis, $v_{s,x'} = \hbar k_{x'}/m_e + 2s\beta_e/\hbar$ is the spin-dependent electron velocity, $f_{\bm k,s}$ is the distribution function in the spin subbands, $\varepsilon_{\bm k,s} = \hbar^2 \bm k^2 /(2 m_e) + 2s\beta_e k_{x'}$ is the electron energy.
To the first order in the spin-orbit splitting, the photocurrent distribution just after the light pulses has the form
\begin{equation}\label{current}
j_{x'}(\varepsilon) =-\frac{2e}{\hbar}\, \frac{m_e \beta_{e}+ m_h \beta_{h}}{m_e+m_h} \frac{d s_{z'}}{d \varepsilon} \varepsilon \:,
\end{equation}
where  $s_{z'}(\varepsilon)$ is the energy distribution of electron spin calculated neglecting the spin-orbit splitting, see Eq.~\eqref{Sk}.
The total photocurrent in the electron subband is given by
\begin{equation}\label{totalcurrent}
J_{x'} = \int\limits_0^\infty j_{x'} (\varepsilon) d \varepsilon = \frac{2e}{\hbar}\, \frac{m_e \beta_{e}+ m_h \beta_{h}}{m_e+m_h}\,  S_{z'} \:,
\end{equation}
where $S_{z'}=\int_0^\infty s_{z'} (\varepsilon) d\varepsilon$ is the total spin density. 
\begin{figure}[t]
\includegraphics[width=0.95\columnwidth]{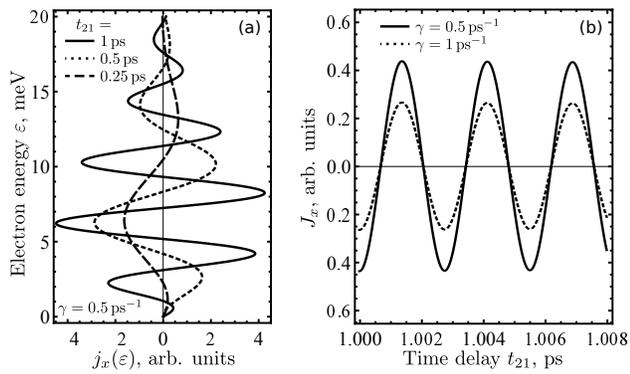}
\caption{(a) Energy distribution of the electric current density induced in the $e1$ subband of (110)-grown QW by two cross-polarized pulses. 
(b) Total electric current as a function of the time delay $t_{21}$ between the pulses. The values of $t_{21}$ and $\gamma$ are indicated on the figures, other parameters of the light pulses are given in the caption of Fig.~\ref{fig2}.}
\label{fig4}
\end{figure}

Figure~\ref{fig4}a shows the energy distribution of the current density $j_{x'} (\varepsilon)$ excited in the $(110)$-grown QWs by cross-polarized light pulses. Similarly to the energy distribution of electron spin, it exhibits an oscillating behavior. The energy profile of the current density depends on the pulse characteristics. The total current in the $e1$ subband as a function of the time delay between the pulses in shown in Fig.~3b. 
Being caused by the quantum interference, the photocurrent is highly sensitive to the phase shift $\Phi_{\bm k}=\omega_{\bm k} t_{21}-\varphi_{21}$. Therefore the change in the delay time $t_{21}$ between the pulses by $\delta t_{21} \sim \pi /\omega \sim 1\,$fs, for the actual parameters, reverses the current direction.  Besides the electric current in the $e1$ subband, the two-pulse optical excitation creates an asymmetric distribution of holes and a photocurrent in the $hh1$ subband as well. In our model, the electron and hole contributions to the total electric current immediately after the optical excitation are equal in strength and opposite in direction. However, due to different momentum relaxation times of electrons and holes, the current contributions decay at different rates giving rise to a total electric current, which is detected in conventional electric measurements. The time-resolved experimental study of the photogalvanic effect enables one to discriminate the electron and hole contributions to the electric current and extract the momentum relaxation times.

To summarize, we have presented the microscopic theory of the optical orientation and photogalvanic effect in quantum wells induced by coherent linearly polarized pulses. The effects can be used to study the interband coherence and the kinetics of charge carriers by optical and electric means.

The work was supported by the Russian Foundation for Basic Research, RF President grants NSh-1085.2014.2 and MD-3098.2014.2, EU projects POLAPHEN and SPANGL4Q, and the ``Dynasty'' Foundation.


\begin{thebibliography}{26}%
\makeatletter
\providecommand \@ifxundefined [1]{%
 \@ifx{#1\undefined}
}%
\providecommand \@ifnum [1]{%
 \ifnum #1\expandafter \@firstoftwo
 \else \expandafter \@secondoftwo
 \fi
}%
\providecommand \@ifx [1]{%
 \ifx #1\expandafter \@firstoftwo
 \else \expandafter \@secondoftwo
 \fi
}%
\providecommand \natexlab [1]{#1}%
\providecommand \enquote  [1]{``#1''}%
\providecommand \bibnamefont  [1]{#1}%
\providecommand \bibfnamefont [1]{#1}%
\providecommand \citenamefont [1]{#1}%
\providecommand \href@noop [0]{\@secondoftwo}%
\providecommand \href [0]{\begingroup \@sanitize@url \@href}%
\providecommand \@href[1]{\@@startlink{#1}\@@href}%
\providecommand \@@href[1]{\endgroup#1\@@endlink}%
\providecommand \@sanitize@url [0]{\catcode `\\12\catcode `\$12\catcode
  `\&12\catcode `\#12\catcode `\^12\catcode `\_12\catcode `\%12\relax}%
\providecommand \@@startlink[1]{}%
\providecommand \@@endlink[0]{}%
\providecommand \url  [0]{\begingroup\@sanitize@url \@url }%
\providecommand \@url [1]{\endgroup\@href {#1}{\urlprefix }}%
\providecommand \urlprefix  [0]{URL }%
\providecommand \Eprint [0]{\href }%
\providecommand \doibase [0]{http://dx.doi.org/}%
\providecommand \selectlanguage [0]{\@gobble}%
\providecommand \bibinfo  [0]{\@secondoftwo}%
\providecommand \bibfield  [0]{\@secondoftwo}%
\providecommand \translation [1]{[#1]}%
\providecommand \BibitemOpen [0]{}%
\providecommand \bibitemStop [0]{}%
\providecommand \bibitemNoStop [0]{.\EOS\space}%
\providecommand \EOS [0]{\spacefactor3000\relax}%
\providecommand \BibitemShut  [1]{\csname bibitem#1\endcsname}%
\let\auto@bib@innerbib\@empty
\bibitem [{\citenamefont {Heberle}\ \emph {et~al.}(1995)\citenamefont
  {Heberle}, \citenamefont {Baumberg},\ and\ \citenamefont
  {K\"ohler}}]{Heberle1995}%
  \BibitemOpen
  \bibfield  {author} {\bibinfo {author} {\bibfnamefont {A.~P.}\ \bibnamefont
  {Heberle}}, \bibinfo {author} {\bibfnamefont {J.~J.}\ \bibnamefont
  {Baumberg}}, \ and\ \bibinfo {author} {\bibfnamefont {K.}~\bibnamefont
  {K\"ohler}},\ }\bibfield  {title} {\enquote {\bibinfo {title} {{Ultrafast
  coherent control and destruction of excitons in quantum wells}},}\ }\href
  {\doibase 10.1103/PhysRevLett.75.2598} {\bibfield  {journal} {\bibinfo
  {journal} {Phys. Rev. Lett.}\ }\textbf {\bibinfo {volume} {75}},\ \bibinfo
  {pages} {2598} (\bibinfo {year} {1995})}\BibitemShut {NoStop}%
\bibitem [{\citenamefont {Marie}\ \emph {et~al.}(1997)\citenamefont {Marie},
  \citenamefont {Le~Jeune}, \citenamefont {Amand}, \citenamefont {Brousseau},
  \citenamefont {Barrau}, \citenamefont {Paillard},\ and\ \citenamefont
  {Planel}}]{Marie1997}%
  \BibitemOpen
  \bibfield  {author} {\bibinfo {author} {\bibfnamefont {X.}~\bibnamefont
  {Marie}}, \bibinfo {author} {\bibfnamefont {P.}~\bibnamefont {Le~Jeune}},
  \bibinfo {author} {\bibfnamefont {T.}~\bibnamefont {Amand}}, \bibinfo
  {author} {\bibfnamefont {M.}~\bibnamefont {Brousseau}}, \bibinfo {author}
  {\bibfnamefont {J.}~\bibnamefont {Barrau}}, \bibinfo {author} {\bibfnamefont
  {M.}~\bibnamefont {Paillard}}, \ and\ \bibinfo {author} {\bibfnamefont
  {R.}~\bibnamefont {Planel}},\ }\bibfield  {title} {\enquote {\bibinfo {title}
  {{Coherent control of the optical orientation of excitons in quantum
  wells}},}\ }\href {\doibase 10.1103/PhysRevLett.79.3222} {\bibfield
  {journal} {\bibinfo  {journal} {Phys. Rev. Lett.}\ }\textbf {\bibinfo
  {volume} {79}},\ \bibinfo {pages} {3222} (\bibinfo {year}
  {1997})}\BibitemShut {NoStop}%
\bibitem [{\citenamefont {Gridnev}(2011)}]{Gridnev2011}%
  \BibitemOpen
  \bibfield  {author} {\bibinfo {author} {\bibfnamefont {V.N.}\ \bibnamefont
  {Gridnev}},\ }\bibfield  {title} {\enquote {\bibinfo {title} {{Optical
  control of the coherent dynamics of excitons in a semiconductor quantum
  well}},}\ }\href {\doibase 10.1134/S0021364011030052} {\bibfield  {journal}
  {\bibinfo  {journal} {Pis'ma Zh. Eksp. Teor. Fiz.}\ }\textbf {\bibinfo
  {volume} {93}},\ \bibinfo {pages} {177} (\bibinfo {year} {2011})},\
  \translation{JETP Lett. {\bf 93}, 161 (2011)}\BibitemShut {NoStop}%
\bibitem [{\citenamefont {Versluis}\ \emph {et~al.}(2010)\citenamefont
  {Versluis}, \citenamefont {Kimel}, \citenamefont {Gridnev}, \citenamefont
  {Yakovlev}, \citenamefont {Karczewski}, \citenamefont {Wojtowicz},
  \citenamefont {Kossut}, \citenamefont {Kirilyuk},\ and\ \citenamefont
  {Rasing}}]{Versluis2010}%
  \BibitemOpen
  \bibfield  {author} {\bibinfo {author} {\bibfnamefont {J.~H.}\ \bibnamefont
  {Versluis}}, \bibinfo {author} {\bibfnamefont {A.~V.}\ \bibnamefont {Kimel}},
  \bibinfo {author} {\bibfnamefont {V.~N.}\ \bibnamefont {Gridnev}}, \bibinfo
  {author} {\bibfnamefont {D.~R.}\ \bibnamefont {Yakovlev}}, \bibinfo {author}
  {\bibfnamefont {G.}~\bibnamefont {Karczewski}}, \bibinfo {author}
  {\bibfnamefont {T.}~\bibnamefont {Wojtowicz}}, \bibinfo {author}
  {\bibfnamefont {J.}~\bibnamefont {Kossut}}, \bibinfo {author} {\bibfnamefont
  {A.}~\bibnamefont {Kirilyuk}}, \ and\ \bibinfo {author} {\bibfnamefont {Th.}\
  \bibnamefont {Rasing}},\ }\bibfield  {title} {\enquote {\bibinfo {title}
  {{Coherence-mediated laser control of exciton and trion spins in CdTe/CdMgTe
  quantum wells studied by the magneto-optical Kerr effect}},}\ }\href
  {http://stacks.iop.org/0953-8984/22/i=11/a=115801} {\bibfield  {journal}
  {\bibinfo  {journal} {J. Phys.: Condens. Matter}\ }\textbf {\bibinfo {volume}
  {22}},\ \bibinfo {pages} {115801} (\bibinfo {year} {2010})}\BibitemShut
  {NoStop}%
\bibitem [{\citenamefont {Langer}\ \emph {et~al.}(2012)\citenamefont {Langer},
  \citenamefont {Poltavtsev}, \citenamefont {Yugova}, \citenamefont {Yakovlev},
  \citenamefont {Karczewski}, \citenamefont {Wojtowicz}, \citenamefont
  {Kossut}, \citenamefont {Akimov},\ and\ \citenamefont {Bayer}}]{Langer2012}%
  \BibitemOpen
  \bibfield  {author} {\bibinfo {author} {\bibfnamefont {L.}~\bibnamefont
  {Langer}}, \bibinfo {author} {\bibfnamefont {S.~V.}\ \bibnamefont
  {Poltavtsev}}, \bibinfo {author} {\bibfnamefont {I.~A.}\ \bibnamefont
  {Yugova}}, \bibinfo {author} {\bibfnamefont {D.~R.}\ \bibnamefont
  {Yakovlev}}, \bibinfo {author} {\bibfnamefont {G.}~\bibnamefont
  {Karczewski}}, \bibinfo {author} {\bibfnamefont {T.}~\bibnamefont
  {Wojtowicz}}, \bibinfo {author} {\bibfnamefont {J.}~\bibnamefont {Kossut}},
  \bibinfo {author} {\bibfnamefont {I.~A.}\ \bibnamefont {Akimov}}, \ and\
  \bibinfo {author} {\bibfnamefont {M.}~\bibnamefont {Bayer}},\ }\bibfield
  {title} {\enquote {\bibinfo {title} {{Magnetic-field control of photon echo
  from the electron-trion system in a CdTe quantum well: Shuffling coherence
  between optically accessible and inaccessible states}},}\ }\href {\doibase
  10.1103/PhysRevLett.109.157403} {\bibfield  {journal} {\bibinfo  {journal}
  {Phys. Rev. Lett.}\ }\textbf {\bibinfo {volume} {109}},\ \bibinfo {pages}
  {157403} (\bibinfo {year} {2012})}\BibitemShut {NoStop}%
\bibitem [{\citenamefont {Toda}\ \emph {et~al.}(2000)\citenamefont {Toda},
  \citenamefont {Sugimoto}, \citenamefont {Nishioka},\ and\ \citenamefont
  {Arakawa}}]{Toda2000}%
  \BibitemOpen
  \bibfield  {author} {\bibinfo {author} {\bibfnamefont {Y.}~\bibnamefont
  {Toda}}, \bibinfo {author} {\bibfnamefont {T.}~\bibnamefont {Sugimoto}},
  \bibinfo {author} {\bibfnamefont {M.}~\bibnamefont {Nishioka}}, \ and\
  \bibinfo {author} {\bibfnamefont {Y.}~\bibnamefont {Arakawa}},\ }\bibfield
  {title} {\enquote {\bibinfo {title} {{Near-field coherent excitation
  spectroscopy of InGaAs/GaAs self-assembled quantum dots}},}\ }\href {\doibase
  http://dx.doi.org/10.1063/1.126810} {\bibfield  {journal} {\bibinfo
  {journal} {Appl. Phys. Lett.}\ }\textbf {\bibinfo {volume} {76}},\ \bibinfo
  {pages} {3887} (\bibinfo {year} {2000})}\BibitemShut {NoStop}%
\bibitem [{\citenamefont {Htoon}\ \emph {et~al.}(2002)\citenamefont {Htoon},
  \citenamefont {Takagahara}, \citenamefont {Kulik}, \citenamefont {Baklenov},
  \citenamefont {Holmes},\ and\ \citenamefont {Shih}}]{Htoon2002}%
  \BibitemOpen
  \bibfield  {author} {\bibinfo {author} {\bibfnamefont {H.}~\bibnamefont
  {Htoon}}, \bibinfo {author} {\bibfnamefont {T.}~\bibnamefont {Takagahara}},
  \bibinfo {author} {\bibfnamefont {D.}~\bibnamefont {Kulik}}, \bibinfo
  {author} {\bibfnamefont {O.}~\bibnamefont {Baklenov}}, \bibinfo {author}
  {\bibfnamefont {A.~L.}\ \bibnamefont {Holmes}}, \ and\ \bibinfo {author}
  {\bibfnamefont {C.~K.}\ \bibnamefont {Shih}},\ }\bibfield  {title} {\enquote
  {\bibinfo {title} {{Interplay of Rabi oscillations and quantum interference
  in semiconductor quantum dots}},}\ }\href {\doibase
  10.1103/PhysRevLett.88.087401} {\bibfield  {journal} {\bibinfo  {journal}
  {Phys. Rev. Lett.}\ }\textbf {\bibinfo {volume} {88}},\ \bibinfo {pages}
  {087401} (\bibinfo {year} {2002})}\BibitemShut {NoStop}%
\bibitem [{\citenamefont {Godden}\ \emph {et~al.}(2012)\citenamefont {Godden},
  \citenamefont {Quilter}, \citenamefont {Ramsay}, \citenamefont {Wu},
  \citenamefont {Brereton}, \citenamefont {Boyle}, \citenamefont {Luxmoore},
  \citenamefont {Puebla-Nunez}, \citenamefont {Fox},\ and\ \citenamefont
  {Skolnick}}]{Godden2012}%
  \BibitemOpen
  \bibfield  {author} {\bibinfo {author} {\bibfnamefont {T.~M.}\ \bibnamefont
  {Godden}}, \bibinfo {author} {\bibfnamefont {J.~H.}\ \bibnamefont {Quilter}},
  \bibinfo {author} {\bibfnamefont {A.~J.}\ \bibnamefont {Ramsay}}, \bibinfo
  {author} {\bibfnamefont {Yanwen}\ \bibnamefont {Wu}}, \bibinfo {author}
  {\bibfnamefont {P.}~\bibnamefont {Brereton}}, \bibinfo {author}
  {\bibfnamefont {S.~J.}\ \bibnamefont {Boyle}}, \bibinfo {author}
  {\bibfnamefont {I.~J.}\ \bibnamefont {Luxmoore}}, \bibinfo {author}
  {\bibfnamefont {J.}~\bibnamefont {Puebla-Nunez}}, \bibinfo {author}
  {\bibfnamefont {A.~M.}\ \bibnamefont {Fox}}, \ and\ \bibinfo {author}
  {\bibfnamefont {M.~S.}\ \bibnamefont {Skolnick}},\ }\bibfield  {title}
  {\enquote {\bibinfo {title} {{Coherent optical control of the spin of a
  single hole in an InAs/GaAs quantum dot}},}\ }\href {\doibase
  10.1103/PhysRevLett.108.017402} {\bibfield  {journal} {\bibinfo  {journal}
  {Phys. Rev. Lett.}\ }\textbf {\bibinfo {volume} {108}},\ \bibinfo {pages}
  {017402} (\bibinfo {year} {2012})}\BibitemShut {NoStop}%
\bibitem [{\citenamefont {Renucci}\ \emph {et~al.}(2003)\citenamefont
  {Renucci}, \citenamefont {Amand},\ and\ \citenamefont {Marie}}]{Renucci2003}%
  \BibitemOpen
  \bibfield  {author} {\bibinfo {author} {\bibfnamefont {P.}~\bibnamefont
  {Renucci}}, \bibinfo {author} {\bibfnamefont {T.}~\bibnamefont {Amand}}, \
  and\ \bibinfo {author} {\bibfnamefont {X.}~\bibnamefont {Marie}},\ }\bibfield
   {title} {\enquote {\bibinfo {title} {{Coherent spin dynamics of polaritons
  in semiconductor microcavities}},}\ }\href
  {http://stacks.iop.org/0268-1242/18/i=10/a=310} {\bibfield  {journal}
  {\bibinfo  {journal} {Semicond. Sci. Technol.}\ }\textbf {\bibinfo {volume}
  {18}},\ \bibinfo {pages} {S361} (\bibinfo {year} {2003})}\BibitemShut
  {NoStop}%
\bibitem [{\citenamefont {Cundiff}(2008)}]{Cundiff2008}%
  \BibitemOpen
  \bibfield  {author} {\bibinfo {author} {\bibfnamefont {S.~T.}\ \bibnamefont
  {Cundiff}},\ }\bibfield  {title} {\enquote {\bibinfo {title} {{Coherent
  spectroscopy of semiconductors}},}\ }\href {\doibase 10.1364/OE.16.004639}
  {\bibfield  {journal} {\bibinfo  {journal} {Opt. Express}\ }\textbf {\bibinfo
  {volume} {16}},\ \bibinfo {pages} {4639} (\bibinfo {year}
  {2008})}\BibitemShut {NoStop}%
\bibitem [{\citenamefont {Ramsay}(2010)}]{Ramsay2010}%
  \BibitemOpen
  \bibfield  {author} {\bibinfo {author} {\bibfnamefont {A.~J.}\ \bibnamefont
  {Ramsay}},\ }\bibfield  {title} {\enquote {\bibinfo {title} {{A review of the
  coherent optical control of the exciton and spin states of semiconductor
  quantum dots}},}\ }\href {http://stacks.iop.org/0268-1242/25/i=10/a=103001}
  {\bibfield  {journal} {\bibinfo  {journal} {Semicond. Sci. Technol.}\
  }\textbf {\bibinfo {volume} {25}},\ \bibinfo {pages} {103001} (\bibinfo
  {year} {2010})}\BibitemShut {NoStop}%
\bibitem [{\citenamefont {Priyadarshi}\ \emph {et~al.}(2012)\citenamefont
  {Priyadarshi}, \citenamefont {Pierz},\ and\ \citenamefont
  {Bieler}}]{Priyadarshi2012}%
  \BibitemOpen
  \bibfield  {author} {\bibinfo {author} {\bibfnamefont {Sh.}\ \bibnamefont
  {Priyadarshi}}, \bibinfo {author} {\bibfnamefont {K.}~\bibnamefont {Pierz}},
  \ and\ \bibinfo {author} {\bibfnamefont {M.}~\bibnamefont {Bieler}},\
  }\bibfield  {title} {\enquote {\bibinfo {title} {{All-optically induced
  ultrafast photocurrents: Beyond the instantaneous coherent response}},}\
  }\href {\doibase 10.1103/PhysRevLett.109.216601} {\bibfield  {journal}
  {\bibinfo  {journal} {Phys. Rev. Lett.}\ }\textbf {\bibinfo {volume} {109}},\
  \bibinfo {pages} {216601} (\bibinfo {year} {2012})}\BibitemShut {NoStop}%
\bibitem [{\citenamefont {Priyadarshi}\ \emph {et~al.}(2013)\citenamefont
  {Priyadarshi}, \citenamefont {Pierz},\ and\ \citenamefont
  {Bieler}}]{Priyadarshi2013}%
  \BibitemOpen
  \bibfield  {author} {\bibinfo {author} {\bibfnamefont {Sh.}\ \bibnamefont
  {Priyadarshi}}, \bibinfo {author} {\bibfnamefont {K.}~\bibnamefont {Pierz}},
  \ and\ \bibinfo {author} {\bibfnamefont {M.}~\bibnamefont {Bieler}},\
  }\bibfield  {title} {\enquote {\bibinfo {title} {All-optically induced
  currents resulting from frequency-modulated coherent polarization},}\ }\href
  {\doibase http://dx.doi.org/10.1063/1.4795722} {\bibfield  {journal}
  {\bibinfo  {journal} {Appl. Phys. Lett.}\ }\textbf {\bibinfo {volume}
  {102}},\ \bibinfo {eid} {112102} (\bibinfo {year} {2013})}\BibitemShut
  {NoStop}%
\bibitem [{\citenamefont {Ivchenko}(2005)}]{Ivchenko_book}%
  \BibitemOpen
  \bibfield  {author} {\bibinfo {author} {\bibfnamefont {E.~L.}\ \bibnamefont
  {Ivchenko}},\ }\href@noop {} {\emph {\bibinfo {title} {{O}ptical spectroscopy
  of semiconductor nanostructures}}}\ (\bibinfo  {publisher} {Alpha Science
  International},\ \bibinfo {address} {Harrow, UK},\ \bibinfo {year}
  {2005})\BibitemShut {NoStop}%
\bibitem [{\citenamefont {Akhiezer}\ and\ \citenamefont
  {Peletminskii}(1981)}]{Akhiezer_book}%
  \BibitemOpen
  \bibfield  {author} {\bibinfo {author} {\bibfnamefont {A.I.}\ \bibnamefont
  {Akhiezer}}\ and\ \bibinfo {author} {\bibfnamefont {S.V.}\ \bibnamefont
  {Peletminskii}},\ }\href@noop {} {\emph {\bibinfo {title} {{Methods of
  Statistical Physics}}}}\ (\bibinfo  {publisher} {Pergamon Press},\ \bibinfo
  {address} {Oxford, UK},\ \bibinfo {year} {1981})\BibitemShut {NoStop}%
\bibitem [{\citenamefont {Tarasenko}(2005)}]{tarasenko2005}%
  \BibitemOpen
  \bibfield  {author} {\bibinfo {author} {\bibfnamefont {S.~A.}\ \bibnamefont
  {Tarasenko}},\ }\bibfield  {title} {\enquote {\bibinfo {title} {{Optical
  orientation of electron spins by linearly polarized light}},}\ }\href
  {\doibase 10.1103/PhysRevB.72.113302} {\bibfield  {journal} {\bibinfo
  {journal} {Phys. Rev. B}\ }\textbf {\bibinfo {volume} {72}},\ \bibinfo
  {pages} {113302} (\bibinfo {year} {2005})}\BibitemShut {NoStop}%
\bibitem [{\citenamefont {Gorelov}\ \emph {et~al.}(2011)\citenamefont
  {Gorelov}, \citenamefont {Tarasenko},\ and\ \citenamefont
  {Averkiev}}]{Gorelov2011}%
  \BibitemOpen
  \bibfield  {author} {\bibinfo {author} {\bibfnamefont {V.A.}\ \bibnamefont
  {Gorelov}}, \bibinfo {author} {\bibfnamefont {S.A.}\ \bibnamefont
  {Tarasenko}}, \ and\ \bibinfo {author} {\bibfnamefont {N.S.}\ \bibnamefont
  {Averkiev}},\ }\bibfield  {title} {\enquote {\bibinfo {title} {{Spin
  orientation of electrons by unpolarized light pulses in low-symmetry quantum
  wells}},}\ }\href {\doibase 10.1134/S1063776111130164} {\bibfield  {journal}
  {\bibinfo  {journal} {Zh. Eksp. Teor. Fiz.}\ }\textbf {\bibinfo {volume}
  {140}},\ \bibinfo {pages} {1002} (\bibinfo {year} {2011})},\
  \translation{JETP {\bf 113}, 873 (2011)}\BibitemShut {NoStop}%
\bibitem [{\citenamefont {Winkler}(2003)}]{Winkler_book}%
  \BibitemOpen
  \bibfield  {author} {\bibinfo {author} {\bibfnamefont {R.}~\bibnamefont
  {Winkler}},\ }\href@noop {} {\emph {\bibinfo {title} {{Spin-orbit coupling
  effects in two-dimensional electron and hole systems}}}}\ (\bibinfo
  {publisher} {Springer},\ \bibinfo {address} {Berlin},\ \bibinfo {year}
  {2003})\BibitemShut {NoStop}%
\bibitem [{\citenamefont {Rashba}\ and\ \citenamefont
  {Sherman}(1988)}]{Rashba1988}%
  \BibitemOpen
  \bibfield  {author} {\bibinfo {author} {\bibfnamefont {E.~I.}\ \bibnamefont
  {Rashba}}\ and\ \bibinfo {author} {\bibfnamefont {E.~Y.}\ \bibnamefont
  {Sherman}},\ }\bibfield  {title} {\enquote {\bibinfo {title} {{Spin orbital
  band splitting in symmetric quantum wells}},}\ }\href {\doibase
  10.1016/0375-9601(88)90140-5} {\bibfield  {journal} {\bibinfo  {journal}
  {{Phys. Lett. A}}\ }\textbf {\bibinfo {volume} {129}},\ \bibinfo {pages}
  {175} (\bibinfo {year} {1988})}\BibitemShut {NoStop}%
\bibitem [{\citenamefont {Durnev}\ \emph {et~al.}(2014)\citenamefont {Durnev},
  \citenamefont {Glazov},\ and\ \citenamefont {Ivchenko}}]{Durnev2014}%
  \BibitemOpen
  \bibfield  {author} {\bibinfo {author} {\bibfnamefont {M.~V.}\ \bibnamefont
  {Durnev}}, \bibinfo {author} {\bibfnamefont {M.~M.}\ \bibnamefont {Glazov}},
  \ and\ \bibinfo {author} {\bibfnamefont {E.~L.}\ \bibnamefont {Ivchenko}},\
  }\bibfield  {title} {\enquote {\bibinfo {title} {{Spin-orbit splitting of
  valence subbands in semiconductor nanostructures}},}\ }\href {\doibase
  10.1103/PhysRevB.89.075430} {\bibfield  {journal} {\bibinfo  {journal} {Phys.
  Rev. B}\ }\textbf {\bibinfo {volume} {89}},\ \bibinfo {pages} {075430}
  (\bibinfo {year} {2014})}\BibitemShut {NoStop}%
\bibitem [{\citenamefont {Gridnev}(2001)}]{Gridnev01}%
  \BibitemOpen
  \bibfield  {author} {\bibinfo {author} {\bibfnamefont {V.~N.}\ \bibnamefont
  {Gridnev}},\ }\bibfield  {title} {\enquote {\bibinfo {title} {Theory of
  faraday rotation beats in quantum wells with large spin splitting},}\ }\href
  {\doibase 10.1134/1.1427126} {\bibfield  {journal} {\bibinfo  {journal}
  {Pis'ma Zh. Eksp. Teor. Fiz.}\ }\textbf {\bibinfo {volume} {74}},\ \bibinfo
  {pages} {417} (\bibinfo {year} {2001})},\ \translation{JETP Lett. {\bf 74},
  380 (2001)}\BibitemShut {NoStop}%
\bibitem [{\citenamefont {Poshakinskiy}\ and\ \citenamefont
  {Tarasenko}(2011)}]{Poshakinskiy11a}%
  \BibitemOpen
  \bibfield  {author} {\bibinfo {author} {\bibfnamefont {A.~V.}\ \bibnamefont
  {Poshakinskiy}}\ and\ \bibinfo {author} {\bibfnamefont {S.~A.}\ \bibnamefont
  {Tarasenko}},\ }\bibfield  {title} {\enquote {\bibinfo {title} {Electron spin
  dephasing in two-dimensional systems with anisotropic scattering},}\ }\href
  {\doibase 10.1103/PhysRevB.84.155326} {\bibfield  {journal} {\bibinfo
  {journal} {Phys. Rev. B}\ }\textbf {\bibinfo {volume} {84}},\ \bibinfo
  {pages} {155326} (\bibinfo {year} {2011})}\BibitemShut {NoStop}%
\bibitem [{\citenamefont {Ivchenko}\ and\ \citenamefont
  {Tarasenko}(2008)}]{Ivchenko2008}%
  \BibitemOpen
  \bibfield  {author} {\bibinfo {author} {\bibfnamefont {E.~L.}\ \bibnamefont
  {Ivchenko}}\ and\ \bibinfo {author} {\bibfnamefont {S.~A.}\ \bibnamefont
  {Tarasenko}},\ }\bibfield  {title} {\enquote {\bibinfo {title} {{Pure spin
  photocurrents}},}\ }\href {http://stacks.iop.org/0268-1242/23/i=11/a=114007}
  {\bibfield  {journal} {\bibinfo  {journal} {Semicond. Sci. Technol.}\
  }\textbf {\bibinfo {volume} {23}},\ \bibinfo {pages} {114007} (\bibinfo
  {year} {2008})}\BibitemShut {NoStop}%
\bibitem [{\citenamefont {Rioux}\ and\ \citenamefont {Sipe}(2012)}]{Rioux2012}%
  \BibitemOpen
  \bibfield  {author} {\bibinfo {author} {\bibfnamefont {J.}~\bibnamefont
  {Rioux}}\ and\ \bibinfo {author} {\bibfnamefont {J.E.}\ \bibnamefont
  {Sipe}},\ }\bibfield  {title} {\enquote {\bibinfo {title} {{Optical injection
  processes in semiconductors}},}\ }\href {\doibase
  http://dx.doi.org/10.1016/j.physe.2012.07.004} {\bibfield  {journal}
  {\bibinfo  {journal} {{Physica E}}\ }\textbf {\bibinfo {volume} {45}},\
  \bibinfo {pages} {1} (\bibinfo {year} {2012})}\BibitemShut {NoStop}%
\bibitem [{\citenamefont {Ganichev}\ and\ \citenamefont
  {Golub}(2014)}]{Ganichev2014}%
  \BibitemOpen
  \bibfield  {author} {\bibinfo {author} {\bibfnamefont {S.~D.}\ \bibnamefont
  {Ganichev}}\ and\ \bibinfo {author} {\bibfnamefont {L.~E.}\ \bibnamefont
  {Golub}},\ }\bibfield  {title} {\enquote {\bibinfo {title} {{Interplay of
  Rashba/Dresselhaus spin splittings probed by photogalvanic spectroscopy -- A
  review}},}\ }\href {\doibase 10.1002/pssb.201350261} {\bibfield  {journal}
  {\bibinfo  {journal} {Phys. Status Solidi B}\ } (\bibinfo {year} {2014}),\
  10.1002/pssb.201350261}\BibitemShut {NoStop}%
\bibitem [{\citenamefont {Shalygin}\ \emph {et~al.}(2007)\citenamefont
  {Shalygin}, \citenamefont {Diehl}, \citenamefont {Hoffmann}, \citenamefont
  {Danilov}, \citenamefont {Herrle}, \citenamefont {Tarasenko}, \citenamefont
  {Schuh}, \citenamefont {Gerl}, \citenamefont {Wegscheider}, \citenamefont
  {Prettl},\ and\ \citenamefont {Ganichev}}]{Shalygin2007}%
  \BibitemOpen
  \bibfield  {author} {\bibinfo {author} {\bibfnamefont {V.A.}\ \bibnamefont
  {Shalygin}}, \bibinfo {author} {\bibfnamefont {H.}~\bibnamefont {Diehl}},
  \bibinfo {author} {\bibfnamefont {Ch.}\ \bibnamefont {Hoffmann}}, \bibinfo
  {author} {\bibfnamefont {S.N.}\ \bibnamefont {Danilov}}, \bibinfo {author}
  {\bibfnamefont {T.}~\bibnamefont {Herrle}}, \bibinfo {author} {\bibfnamefont
  {S.A.}\ \bibnamefont {Tarasenko}}, \bibinfo {author} {\bibfnamefont
  {D.}~\bibnamefont {Schuh}}, \bibinfo {author} {\bibfnamefont {Ch.}\
  \bibnamefont {Gerl}}, \bibinfo {author} {\bibfnamefont {W.}~\bibnamefont
  {Wegscheider}}, \bibinfo {author} {\bibfnamefont {W.}~\bibnamefont {Prettl}},
  \ and\ \bibinfo {author} {\bibfnamefont {S.D.}\ \bibnamefont {Ganichev}},\
  }\bibfield  {title} {\enquote {\bibinfo {title} {{Spin photocurrents and the
  circular photon drag effect in (110)-grown quantum well structures}},}\
  }\href {\doibase 10.1134/S0021364006220097} {\bibfield  {journal} {\bibinfo
  {journal} {Pis'ma Zh. Eksp. Teor. Fiz.}\ }\textbf {\bibinfo {volume} {84}},\
  \bibinfo {pages} {666} (\bibinfo {year} {2007})},\ \translation{JETP Lett.
  {\bf 84}, 570 (2007)}\BibitemShut {NoStop}%
\end{thebibliography}
\end{document}